\documentclass[letter]{aa}
\usepackage{natbib}
\bibpunct{(}{)}{;}{a}{}{,} % to follow the A&A style
\usepackage{color}
\usepackage{xspace}
\usepackage{amssymb}
\usepackage{amsmath}
\usepackage{graphicx}
\usepackage{balance}

\def\uf{\ensuremath{u_\mathrm{f}}\xspace}

\def\cs{\ensuremath{c_\mathrm{s}}\xspace}

\def\Hp{\ensuremath{H_\mathrm{p}}\xspace}

\def\Sc{\ensuremath{\mathrm{Sc}}\xspace}
\def\St{\ensuremath{\mathrm{St}}\xspace}

\def\dx{\ensuremath{\mathrm{d}}}

\def\inprep{in prep.\xspace}
\newcommand{\del}{\partial}%

\definecolor{darkgreen}{rgb}{0,0.6,0}
\makeatletter
\if@referee \newcommand{\changed}[1]{\textbf{#1}} \else \newcommand{\changed}[1]{#1} \fi 
\makeatother
\usepackage[%
  pdfauthor={T. Birnstiel, C.P. Dullemond, F. Brauer},
  pdftitle={Dust retention in protoplanetary disks},
  linktocpage,
  pdfstartview=FitH,
  breaklinks,
  linkcolor=blue,
  anchorcolor=blue,
  citecolor=blue,
  filecolor=blue,
  menucolor=blue,
  urlcolor=blue,
  colorlinks=true]{hyperref}
\defcitealias{Brauer:2008p215}{B08a}
\begin{document}
\title{Dust retention in protoplanetary disks}
\titlerunning{Dust retention in protoplanetary disks}
\author{T.~Birnstiel \and C.P.~Dullemond \and F.~Brauer}
\authorrunning{T.~Birnstiel~et~al.}
\institute{Junior Research Group at the Max-Planck-Institut f\"ur Astronomie, K\"onigstuhl 17, D-69117 Heidelberg, Germany.\\Email: \texttt{birnstiel@mpia.de}} \date{Received 8 May 2009 / Accepted 3 July 2009}

\abstract{
  Protoplanetary disks are observed to remain dust-rich for
  up to several million years. Theoretical modeling, on the other hand,
  raises several questions. Firstly, dust coagulation occurs so rapidly, that if
  the small dust grains are not replenished by collisional fragmentation of
  dust aggregates, most disks should be observed to be dust poor, which is
  not the case. Secondly, if dust aggregates grow to sizes of the order of
  centimeters to meters, they drift so fast inwards, that they
  are quickly lost.}
  {We attempt to verify if collisional fragmentation
  of dust aggregates is effective enough to keep disks `dusty' by replenishing
  the population of small grains and by preventing excessive radial drift.}
  {With a new and sophisticated implicitly integrated
  coagulation and fragmentation modeling code, we solve the combined problem
  of coagulation, fragmentation, turbulent mixing and radial drift
  and at the same time solve for the 1-D viscous gas disk
  evolution.}
  {We find that for a critical collision velocity of 1~m/s, as
  suggested by laboratory experiments, the fragmentation is so
  effective, that at all times the dust is in the form of relatively small
  particles. This means that radial drift is small and that large amounts
  of small dust particles remain present for a few million years, as
  observed. For a critical velocity of 10~m/s, we find that particles
  grow about two orders of magnitude larger, which leads again to significant
  dust loss since larger particles are more strongly affected by radial drift.}
{}


\maketitle

\section{Introduction}\label{sec:introduction}
The quest for a comprehensive understanding of the formation of planets in
the dusty circumstellar disks around many young stars has gained significant
impetus. This is partly because of developments in the numerical
modeling of these processes, but also because of the high-quality infrared and
\mbox{(sub-)}millimeter data that are now available of hundreds of T Tauri and
Herbig Ae star disks from observatories such as the Spitzer Space Telescope
\citep[e.g.][]{Furlan:2006p4444,KesslerSilacci:2006p4450}, the Very
Large Telescope \citep[e.g.][]{vanBoekel:2004p4708}, the Submillimeter Array
\citep[e.g.][]{Andrews:2007p3380} and the Plateau de Bure Interferometer
\citep[e.g.][]{Pietu:2007p4501}. The challenge for theoretical
astrophysicists is now not only to come up with a plausible model of how to
overcome the various problems in our current understanding of how planets
form, but also to ensure that new models are consistent with the
observational data obtained from these planetary birthplaces.

One of the challenges of theories of planet formation is what is often called the `meter-size barrier'. The first stages of planet formation are believed to involve the coagulation of dust particles to aggregates of ever larger size, and eventually to the size of asteroids or comets \citep{Nakagawa:1981p4533,Weidenschilling:1980p4572,Weidenschilling:1984p4590,Weidenschilling:1997p4593}. It was shown by \citet{Weidenschilling:1977p865} that as dust particles grow by coagulation, they acquire increasingly large relative velocities with respect to other particles in their vicinity, as well as a systematic inward drift velocity as originally described by \citet{Whipple:1972p4621}. The high relative velocities lead to destructive collisions \citep{Blum:2008p1920}, and thereby limit the growth of the aggregates to some maximum size. The radial drift also leads to a loss of solids toward the star well before large bodies can even form \citep[][hereafter \citetalias{Brauer:2008p215}]{Brauer:2008p215}.

\citet{Dullemond:2005p378} showed that if, hypothetically, coagulation can proceed without
limit, i.e.,\ if no fragmentation occurs and no radial drift is present, the
growth is so efficient that within $10^5$ years very few small ($a\lesssim
1$mm) opacity bearing grains remain in the disk. The disk becomes optically
thin and the mid- to far-infrared flux of the disk drops to levels well
below that observed in the majority of gas-rich circumstellar disks
around pre-main sequence stars. By excluding other mechanisms, \citet{Dominik:2008p4626} argue that fragmentation is the most promising process for maintaining a high abundance of fine-grained dust in evolved disks.

However, it has never been demonstrated explicitly that fragmentation is efficient enough to
retain the population of small grains at the required levels. Moreover, if
we include the effect of radial drift, it remains to be seen whether this
radial drift is suppressed sufficiently by keeping the grains small through
fragmentation. It is the goal of this Letter to investigate this issue and to
determine how efficient fragmentation should be to agree with observations.

\section{Model}\label{sec:model}
The model presented here is a combination of a 1D viscous gas disk evolution code and a dust evolution code treating the turbulent mixing, radial drift, coagulation and fragmentation of the dust. A detailed description of the model will follow in a later paper (Birnstiel et al., \inprep). Although the evolution of dust can influence the gas disk evolution, we neglect this effect in our model.

The gas disk model is similar to that described in \citet{Hueso:2005p685}. The mid-plane temperature of the disk is approximated by following \citet{Nakamoto:1994p798}, taking irradiation by the central star and viscous heating into account. The viscous evolution of the disk surface density $\Sigma_\mathrm{g}$ is described by \citep[see][]{LyndenBell:1974p1945},
\begin{equation}
\frac{\del \Sigma_\mathrm{g} }{\del t} - \frac{1}{r}\frac{\del}{\del r}\left( \Sigma_\mathrm{g} \, r \, u_\mathrm{g}\right) = 0,
\end{equation}
where the gas radial velocity $u_\mathrm{g}$ is given by
\begin{equation}
u_\mathrm{g} = - \frac{3}{\Sigma_\mathrm{g}\sqrt{r}} \frac{\del}{\del r} \left( \Sigma_\mathrm{g} \nu_\mathrm{g} \sqrt{r} \right).
\end{equation}
$\nu_\mathrm{g}$ is the gas turbulent viscosity,
\begin{equation}
\nu_\mathrm{g} = \alpha \, \cs \, \Hp,
\end{equation}
\cs denotes the sound speed, \Hp the pressure scale height, and $\alpha$ is the turbulence parameter \citep{Shakura:1973p4854}.

Grains are affected by a systematic radial inward drift related to headwind caused by the pressure-supported gas \citep{Weidenschilling:1977p865},
\begin{equation}
u_\text{drift} = \frac{1}{\St^{-1}+\St}\cdot\frac{ \del_{r}{P_\text{g}}}{\rho_
\text{g} \: \Omega_\text{k}},
\label{eq:v_drift}
\end{equation}
where \St is the Stokes number of the particle (a dimensionless representation of the particle size), $P_\text{g}$ is the gas pressure, $\rho_\text{g}$ is the gas volume density, and $\Omega_\text{k}$ is the Kepler frequency.
In the Epstein regime, \St is given by
\begin{equation}
\St = \frac{a \rho_\mathrm{s}}{\Sigma_\mathrm{g}} \frac{\pi}{2},
\label{eq:stokesnumber}
\end{equation}
where $\rho_\mathrm{s}$ is the solid density of the dust grains.

The second contribution to radial dust velocities is the gas drag due to the radial motion of the gas,
\begin{equation}
u_\text{drag} = \frac{u_\text{g}}{\Sc},
\label{eq:v_drag}
\end{equation}
where $\Sc=1 + \St^2$ is the Schmidt number, following \citet{Youdin:2007p2021}.

The coupling to the gas turbulence leads to turbulent mixing of each dust species with a diffusion constant that is taken to be
\begin{equation}
D_\text{d} = \frac{\nu_\mathrm{g}}{\Sc}.
\end{equation}

The dust grain number density $n(m,r,z)$ is a function of mass $m$, radius $r$, height above the mid-plane $z$, and time. If we define
\begin{equation}
\sigma(m,r) \equiv \int_{-\infty}^{\infty} n(m,r,z) \, m^2 \dx z,
\end{equation}
the vertically integrated time-evolution of this distribution can now be described by a general two-body process and an advection-diffusion equation,
\begin{equation}
\begin{array}{lll}
\frac{\partial \sigma(m,r)}{\partial t} 
&=&\iint_{0}^{\infty} K(m,m',m'') \, \sigma(m',r) \, \sigma(m'',r)\, \dx m'\, \dx m''\\
\\
&&-\frac{1}{r}\frac{\partial}{\partial r}
\left[ r \cdot \sigma(m,r) \cdot (u_\mathrm{drag}+u_\mathrm{drift}) \right]\\
\\
&&+\frac{1}{r}\frac{\partial}{\partial r} \left[ r \cdot D_\mathrm{d} \cdot \frac{\partial}{\partial r} \left( \frac{\sigma(m,r)}{\Sigma_\mathrm{g}}\right) \cdot \Sigma_\mathrm{g} \right].
\end{array}
\label{eq:smolu}
\end{equation}
Here, the right-hand side terms (from top to bottom) correspond to coagulation/fragmentation, advection, and turbulent mixing.

Since the detailed version of the coagulation/fragmentation equation is lengthy
\citetext{\citealp{Weidenschilling:1980p4572}; \citealp{Nakagawa:1981p4533}; \citealp{Dullemond:2005p378}; \citetalias{Brauer:2008p215}}, we include here only the general integral of a two-body process (first term on the right hand side). For a detailed description of the physics of coagulation/fragmentation, we refer to \citetalias{Brauer:2008p215}; the numerical implementation will be discussed in Birnstiel et al. (in prep.).

The physical effects that produce the relative particle velocities considered here are Brownian motion, relative radial motion, vertical settling \citepalias[see][]{Brauer:2008p215}, and turbulent motion \citep[see][]{Ormel:2007p801}.

If particles collide with a relative velocity higher than the critical velocity \uf, they either fragment into a power-law size distribution of fragments \citepalias[i.e., $n(m)\propto m^{-1.83}$, see][]{Brauer:2008p215} or the smaller body excavates mass from the larger one (cratering). We assume that the amount of excavated mass equals the mass of the smaller body. The fragmentation velocity \uf is a free parameter in our model, and unless otherwise noted is taken to be 1~m/s, as suggested by experiments \citep[e.g.,][]{Blum:1993p4324} and by theoretical modeling \citep[e.g.,][]{Stewart:2009p5281}.

We note that the velocities and therefore also the relative velocities are dependent on the particle size. If we assume that relative particle velocities are dominated by turbulent motion,
\begin{equation}
\Delta u_\mathrm{turb} \propto \sqrt{\alpha\:\St} \: \cs,
\label{eq:v_turb}
\end{equation}
then we can estimate the Stokes number of the largest particles by equating the relative velocity and fragmentation velocity,
\begin{equation}
\St_\mathrm{max} \simeq \frac{\uf^2}{\alpha\:\cs^2}.
\label{eq:st_max}
\end{equation}
With Eq. \ref{eq:stokesnumber}, we can relate this maximum `dimensionless size' $\St_\mathrm{max}$ to a particle size,
\begin{equation}
a_\mathrm{max} \simeq \frac{2\Sigma_\mathrm{g} }{\pi \alpha \rho_\mathrm{s}} \cdot \frac{\uf^2}{\cs^2}.
\label{eq:a_max}
\end{equation}
It should be noted that Eq. \ref{eq:v_turb} is an approximation of the equations derived in \citet{Ormel:2007p801}. The detailed relative velocities of particles depend on the Stokes numbers of both particles. Hence, Eq. \ref{eq:a_max} is an order of magnitude estimate.

The combined equations for coagulation, fragmentation, and radial motion due to drift, drag, and mixing are solved using the technique of implicit integration \citepalias[similar to][]{Brauer:2008p215}. We use the SixPACK F-90 linear algebra package\footnote{available from \url{www.engineers.auckland.ac.nz/~snor007}} to solve the sparse matrix equation.

\begin{figure}[t]
	\resizebox{\hsize}{!}{\includegraphics{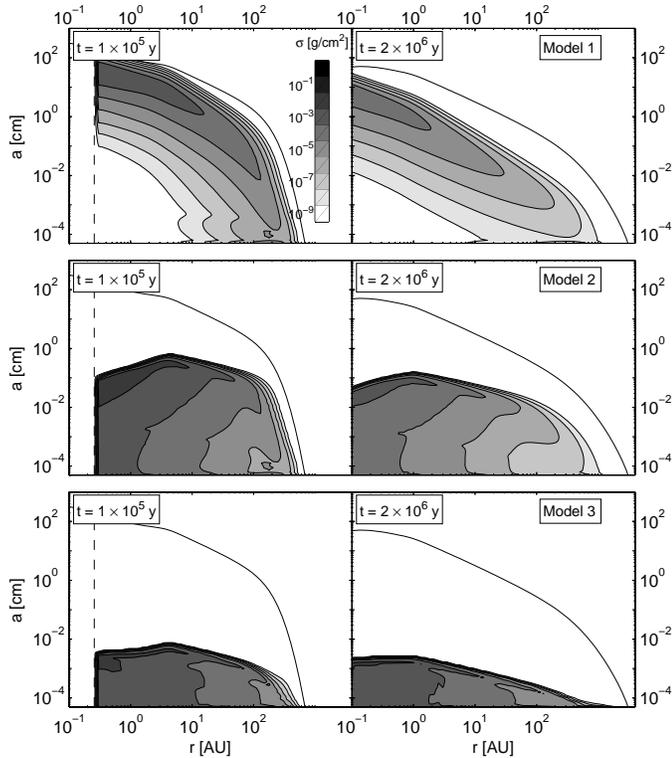}}
   \caption{Snapshots of the vertically integrated dust density distribution of \changed{model}~1 (only coagulation), \changed{model}~2 (fragmentation at 10~m/s) and \changed{model}~3 (fragmentation at 1~m/s). The grain size is given by $a=(3m/4\pi\rho_\mathrm{s})^{1/3}$. The dashed line shows the evaporation radius within which no coagulation is calculated. The solid line shows the particle size corresponding to a Stokes number of unity.}
   \label{fig:snapshots}
\end{figure}

\begin{figure}[t]
   \resizebox{\hsize}{!}{\includegraphics{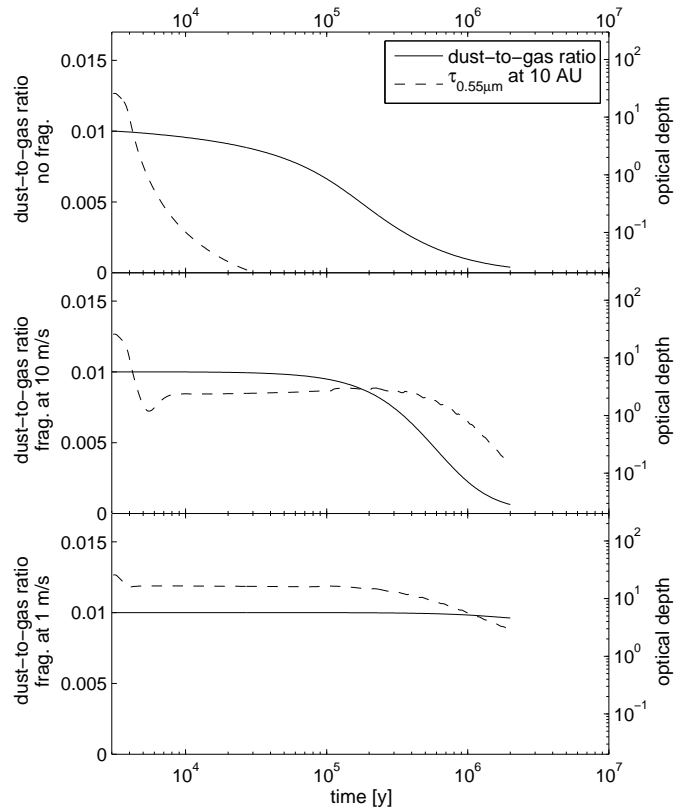}}
   \caption{Dust-to-gas ratio (solid line, linear scale) and 0.55 $\mu$m optical depth at 10 AU (dashed line, logarithmic scale) as function of time for \changed{model}~1 (top), \changed{model}~2 (center) and \changed{model}~3 (bottom).}
   \label{fig:d2g_1}
\end{figure}

\section{Results}\label{sec:results}
The simulation results presented here begin with a 0.07~$M_\odot$ disk of power-law surface density $\Sigma(r)\propto r^{-1}$ (from 0.05 to 150 AU) around a 0.5~$M_\odot$ star. Unless otherwise noted, we used a turbulence parameter of $10^{-2}$. The three models differ in the prescription of fragmentation. In \changed{model}~1, we accounted only for growth without fragmentation. In \changed{model}~2, particles fragment at $\uf=10$ m/s, whereas in \changed{model}~3, the fragmentation speed is taken to be 1~m/s. Selected  results of all three models are depicted in \changed{Figure} \ref{fig:snapshots}.

The top row shows the evolution in the surface density distribution of dust because of coagulation, radial mixing, and radial drift in an evolving protoplanetary disk, corresponding to \changed{model}~1.
The dashed line denotes the grain evaporation radius, which moves inwards as the temperature falls (lower surface density caused by accretion produces less viscous heating).
The evaporation radius in our model is defined as the radius where the temperature rises above 1500~K assuming that all particles evaporate at this temperature. The solid line denotes the dust grain size, which corresponds to a Stokes number of unity.  It can be seen that particles grow to decimeter size and then quickly drift inside the evaporation radius.

The results of simulations that also include grain fragmentation are plotted in the second and third row of \changed{Figure} \ref{fig:snapshots}.
In the innermost regions of model 2, relative particle velocities are high enough (as a result of shorter dynamical times) to fragment particles already at relatively small Stokes number. However, the maximum size of the grain distribution is still strongly affected by radial drift, especially at larger radii. Larger particles are subject to radial drift and \changed{therefore} drift to smaller radii, where they are pulverized.

In contrast, fragmentation becomes the main limiting factor for particle growth throughout the disk if the critical fragmentation velocity is 1~m/s (\changed{model}~3). The result is that particles fragment at a small Stokes number. Particles with a Stokes number less than about $10^{-3}$ are still strongly coupled to the gas and not subject to strong radial drift \citep[see][]{Brauer:2007p232}. Therefore, significant amounts of dust can be retained for several million years if fragmentation is included in the calculations.

\changed{Figure} \ref{fig:d2g_1} compares the evolution in the dust-to-gas ratio (solid line) and the 0.55 $\mu$m optical depth at 10 AU (dashed line) of all three simulations.
The optical depth is defined as
\begin{equation}
\tau_\nu = \int_{0}^\infty \sigma(m) \, \kappa_\nu(m) \dx m.
\end{equation}
It can be seen that the dust-to-gas ratio declines dramatically in the first two cases. The optical depth in \changed{model}~1 drops on even shorter timescales than the dust-to-gas ratio because not only the dust loss but also the dust growth causes the optical depth to decrease.

\changed{model}~2 can retain the optical depth longer because small particles remain (due to fragmentation), providing enough surface to keep the disk optically thick for about 0.9 Myr. The initial dip in the optical depth is caused by the initial condition: since initially, all mass is in small grains, the optical depth decreases as particles grow to larger sizes until fragmentation sets in, which increases the optical depth until a semi-equilibrium of growth and fragmentation is reached.

If particles are already fragmented at 1 m/s (\changed{model}~3), the dust-to-gas ratio declines only slightly, after 2 My, only about 4\% of the dust being lost (compared to 96\% and 94\% in models~1 and 2). The optical depth decreases on longer timescales than in \changed{model}~2 since the dust mass is lost only on the viscous timescale.

This mechanism of grain retention relies on the particles remaining small due to fragmentation. From Eq. \ref{eq:st_max} and \ref{eq:a_max}, it follows that the maximum size or Stokes number depends strongly on the critical velocity \uf and less strongly on $\alpha$. If \uf is 10 m/s instead of 1 m/s, the corresponding particle size is 100 times larger meaning that the largest particles are now also experiencing the radial drift barrier and therefore drifting quickly towards the central star.

\section{Discussion and conclusion}\label{sec:discussion}
We have shown that grain fragmentation plays an important role, not only in keeping the disk optically thick (by replenishing small grains) but also in maintaining a relatively high abundance of solids of all sizes.
If particles fragment at relatively small sizes, they can stay below the size-regime where they are affected by radial drift.

The simulation results presented here suggest that fragmentation of relatively low critical velocity (a few m/s) is needed to retain the dust required for planet formation, although the formation of planetesimals probably requires periods (and/or regions) of quiescence such as dead zones or pressure bumps \citep[see][]{Kretke:2007p697,Brauer:2008p212}. These locally confined quiescent environments could decrease both the relative and radial velocities of particles, thus allowing particles possibly to either grow to larger sizes or to accumulate until gravitational instabilities become important \citep[see][]{Johansen:2007p4788,Lyra:2009p4812}. These effects are not yet included in this model and will be the subject of future research.

Assuming that the proposed mechanism of dust retention is effective in protoplanetary disks, we can constrain the critical fragmentation velocity to be smaller than 10 m/s since velocities $\gtrsim$ 10 m/s cannot explain the observed lifetimes of disks even in the case of small $\alpha$.

\begin{acknowledgements}
We like to thank Thomas Henning, Chris Ormel and Andras Zsom for useful discussions.
We also wish to thank the anonymous referee for his fast and insightful report which helped to improve the paper.
\end{acknowledgements}
\balance
\bibliographystyle{aa}
\bibliography{bibliography}
\end{document}